\documentclass[aps,twocolumn]{revtex4}

\usepackage{amsmath}
\usepackage{amsfonts}
\usepackage{amssymb}

\usepackage{graphics}
\usepackage{amsmath}
\usepackage{amsfonts}
\usepackage{amssymb}
\usepackage{graphicx}
\usepackage{epsfig}

\def\ba{\begin{eqnarray}}
\def\ea{\end{eqnarray}}
\def\be{\begin{equation}}
\def\ee{\end{equation}}

\def\d{\mathrm{d}}

\def\mn{_{\mu \nu}}
\def\mupn{^\mu_{\, \nu}}
\def\({\left(}
\def\){\right)}

\def\kf{\kappa_5^2}
\def\kt{\tilde{\kappa}^2}
\def\l{\ell}

\begin{document}

\title{Weaker Gravity at Submillimetre Scales in Braneworld Models}

\author{Claudia de Rham$^{(1)}$, Tetsuya Shiromizu$^{(2)}$ and Andrew J. Tolley$^{(3)}$}
\affiliation{$^{(1)}$Ernest Rutherford Physics Building, McGill University,
Montreal, QC H3A 2T8, Canada.}
\affiliation{$^{(2)}$Department of Physics, Tokyo Institute of
Technology, Tokyo 152-8551, Japan}
\affiliation{$^{(3)}$Joseph Henry
Laboratories, Princeton University, Princeton NJ, 08544, USA.}

\date{\today}

\begin{abstract}
Braneworld models typically predict gravity to grow stronger at short
distances. In this paper, we consider braneworlds with two types of
additional curvature couplings, a Gauss-Bonnet (GB) term in the bulk,
and an Einstein-Hilbert (EH) term on the brane. In the regime where
these terms are dominant over the bulk EH term, linearized gravity
becomes weaker at short distances on the brane.
In both models, the weakening of gravity is tied to the presence of ghosts in the graviton mass spectrum.
We find that the ordinary coupling of matter to
gravity is recovered at low energies/long wavelengths on the brane.
We give some implications for cosmology and show its
compatibility with observations.
We also discuss the stability of compact stars.
\end{abstract}
\maketitle

\section{Introduction}

Current measurements of gravity do not deviate from Newton's law at distances greater than $0.2$mm
\cite{Hoyle:2000cv}, leaving open the possibility that gravity might be modified at shorter
length scales.
The potential observation that gravity is weaker at small distances
would be a serious challenge to current high energy theories, forcing
us to develop unconventional ideas. In this spirit, we shall consider
a class of examples framed within braneworld models where gravity  can
be weaker than Newton's law at small distances.  For a review on
theoretical models predicting a deviation from Newton's inverse-square law,
and on experimental tests, see Ref. \cite{Adelberger:2003zx}.

In braneworld scenarios, gravity on the brane is expected to
be modified at distances of order of the size of the
extra-dimension \cite{ADD}, or the bulk curvature scale
\cite{Randall:1999vf}. The measurement of gravity at submillimeter scales
might therefore provide a way to test braneworld scenarios.
In particular, the Randall-Sundrum
(RS) model \cite{Randall:1999vf}, gives rise to a modification of gravity which
appears stronger at short distances \cite{Garriga:1999yh,Callin:2004py}.
This is due to the presence of massive Kaluza-Klein (KK) modes which
contribute negatively to the gravitational potential on the brane. In any braneworld model, the presence
of the KK modes of the graviton will typically modify the gravitational potential generated by a
source of mass $M$ in the following way
\ba
V(r)=-MG_4\(\frac{1}{r}+\int_0^{\infty} \d m \,  \frac{e^{-m r}}{r} \rho(m)\),
\ea
a result which follows from the Kallen-Lehmann spectral representation of the propagator.
Here $\rho(m)$ measures the influence of the KK modes of mass $m$ on
the brane \footnote{If the extra-dimension is compactified, the KK modes are
discrete and the integral should be understood as a sum over
the different modes.}. For gravity to appear weaker in any regime,
the spectral density $\rho(m)$ should be negative for some range of
$m$. However, unitarity (absence of ghosts) usually requires $\rho(m)
\ge 0$.

This result is generic to many modifications of gravity, for instance
if there are additional gravitational strength interactions mediated
by other massive spin-0 or spin-2 particles, (see for instance Ref. \cite{Callin:2005wi}). However, there are
exceptions, for instance it has been proposed that the graviton is a
composite particle, {\it ie.} has some finite size
\cite{Sundrum:2003jq,Sundrum:2003tb}, and that the force of gravity
would be naturally cut of at this size. In Ref. \cite{Biswas:2005qr},
higher derivative modifications to the gravitational action were
considered making it `asymptotically safe', also providing a cutoff to
the force of gravity at small distances.
A less radical proposal would be an additional weakening from the exchange of massive spin-1
bosons which would give a repulsive Yukawa modification to Newton's law
\cite{Adelberger:2003zx}. These new gauge fields could be propagating
in the bulk of any typical braneworld scenario, and could mediate a
force of the same order of magnitude as gravity. Such gauge
fields would necessarily violate the Equivalence Principle and
the gauge coupling to ordinary matter would be strongly
constrained by experiments \cite{Smith:1999cr}.

In the examples that follow we shall consider the
possibility that $\rho(m)<0$, signalling the presence of
ghosts. Ghosts in a theory are typically a problem for two reasons:
Firstly they suggest the presence of a classical instability, the
significance of which needs to be addressed in the context of a
specific model.
Secondly they signal the breakdown of unitarity, or of a quantum
instability due to their carrying negative energy, see Ref. \cite{Cline:2003gs}.
However, if conventional gravity is recovered at low energies, as will
be the case in our examples, unitarity is expected to be restored in
that limit.
In what follows we
shall simply assume that the usual notions of quantum field theory do
not apply to gravitation at least at submillimetre scales, and that
some unusual notion of quantum gravity resolves this issue  \cite{Hawking:2001yt}.

In this work, we consider the potential corrections arising
on a brane embedded in a five-dimensional space-time. Although the
preceding argument suggests the presence of instabilities, we show how these models are nonetheless
classically viable, and are consistent with cosmological
observations.

As a first example, we consider the extension of the RS model
to the case when GB terms
are present in the bulk \cite{Wheeler:1985qd,Neupane:2001st,Deruelle:2003tz,Maeda:2003vq,Neupane:2001kd,Cho:2001nf,Bronnikov:2006jy}.
In this work, we explore the
possibility that these terms have a contribution of the same order of
magnitude as the EH term. Such a regime is not usually taken into
account in the literature, where the GB terms are usually thought to
arise from quantum/string corrections (see Ref. \cite{Boulware:1985wk}) and are therefore small.
In a five-dimensional spacetime, the GB terms are
the unique functions of the metric that do not
alter the Cauchy problem \cite{Meissner:2000dy}. This implies that
there are no new degrees of freedom and the equations of motion have
no higher than second order time derivatives
\cite{Deruelle:2003ck,Kakushadze:2001bd}.
 Once the GB terms are important, nothing allows us
a priori to ignore the higher-order corrections terms.
We shall however assume a large hierarchy between the GB term and
other higher order corrections. Hierarchy problems are common to
gravity, naively
one would expect the cosmological constant to be of order of magnitude
$10^{120}$ times more than its actual observed value. It is not clear
then that our usual notions of naturalness apply to gravity, and we
will therefore explore this possibility in what follows.

We find in this example that the contribution from
the KK corrections actually reverses sign, leading to weaker gravity
at short distances. The expected instability associated with this
mechanism can be understood when we study the response to bulk
matter. In this regime, the gravity part in the five-dimensional
action reverses sign, leading to a ``wrong'' coupling between
gravity and bulk matter. There is however no instability associated
with matter on the brane, and the theory on the brane is stable, at
least at the classical level.

Motivated by this result, we consider an alternative scenario,
where the gravitational response to matter on the bulk will be
stable. We introduce a negative EH term into the brane
action and imagine that matter on the brane couples with the ``wrong sign'' action.
In that case gravity will respond the correct way to bulk
matter far away from the brane, and the correct way to any matter
confined on the brane. The theory is therefore stable, at
least classically and the theory on the brane presents all the general
features necessary to explain the cosmological behaviour of our
Universe. Gravity will however appear weaker on short scales.

This work is organized as follows.
We start by reviewing the RS model in presence of GB terms in section
\ref{sec2}. After presenting the background spacetime behaviour, we study the
response of gravity to a static source on the brane and show that
gravity appears weaker at short-distances. We then discuss the
implications for cosmology, and show that at low-energy the
four-dimensional behaviour is recovered. The stability of this model
is then studied and the response to bulk matter is shown to be
unstable.
We then focus on an alternative model in section \ref{sec alt}, where
no GB terms are present in the bulk, but instead a EH term on the
brane action which we take with a negative sign. The consequences for
cosmology are explored and the gravity is shown to be weaker at short
scales. We show that both the weak and null energy conditions are
satisfied for the effective energy density provided it is
satisfied by the matter field confined on the brane.
Finally we present other physical implications of these two models in
section \ref{secOI}. In particular we discuss the implications of
weaker gravity on the stability
of massive stars. For weaker gravity at short distances, we show that
the stability of compact stars is improved.

In what follows, we use the index notation that Roman indices are
fully five-dimensional, while Greek indices are four dimensional,
labeling the transverse $x^\mu$ direction along the brane.
Roman fonts $\mathrm{R}, \mathrm{G}$, designate quantities with
respect to the five-dimensional metric, whereas normal fonts $R,G$
designate quantities evaluated with respect to the induced metric on
the brane.

\section{Weaker gravity from GB braneworld model}
\label{sec2}

\subsection{GB braneworld model}

Our starting point is the five-dimensional action
\ba
S^{(\text{5d})}=\frac{1}{2 \kf}\int \d ^5x\sqrt{-g} \left[
\mathrm{R}-2\Lambda+\frac{\beta  \l^2}{4}\, \mathcal{R}_2
\right]+S^{(\text{4d})}_b, \label{5daction}
\ea
where $\Lambda$ is
the bulk cosmological constant. In general $\Lambda$ is considered
to be negative so that the bulk vacuum geometry is Anti-de Sitter (AdS),
but as we shall see this condition can be relaxed for some values of
the coefficient $\beta$ of the GB terms $\mathcal{R}_2$. The AdS
length scale of our chosen vacuum is denoted as $\l$, and the GB term $\mathcal{R}_{2}$
is the trace of
\ba
 \mathcal{R}^{\,A}_{2\, B}&=&\mathrm{R} \, \mathrm{R}^A_B-2
 \,\mathrm{R}^A_ C\,\mathrm{R}^C_B-2  \mathrm{R}^{CD}\,\mathrm{R}^A_{\ \
   CBD}\\
&& +\mathrm{R}^A_{\ \ DEF}\, \mathrm{R}_B^{\ \ DEF} ,\notag
\ea
where $\mathrm{R}^A_{\ \ BCD}$ is the five-dimensional Riemann tensor.
Finally the term $S^{(\text{4d})}_b$ in \eqref{5daction} is the four dimensional action for the brane:
\ba
S^{(\text{4d})}_b=\int_{\text{brane}}\d^4 x\sqrt{-q}\left(Q+\mathcal{L}_m-\lambda \right),
\ea
where $Q$ gives us the generalization of the
Gibbon-Hawking's boundary terms \cite{Gibbons:1976ue,Myers:1987yn}:
\ba
Q=2K+\beta \l^2 \(J-2 G^\mu_\nu K^\nu_\mu\),
\ea
$K\mupn$ being the extrinsic curvature on the brane, $G\mupn$
being the Einstein tensor on the brane and
\ba
J\mupn=-\frac 2 3 K^\mu_\alpha K^\alpha_\beta K^\beta_\nu+\frac 2 3 K\,
K^\mu_\alpha K^\alpha_\nu+\frac 1 3 K\mupn (K^\alpha_\beta K^\beta_\alpha-K^2).
\ea
In the brane action, we separate the brane tension $\lambda$ from
the brane matter fields Lagrangian $\mathcal{L}_m$.

In the five-dimensional bulk, the Einstein equations are
\ba
\mathcal{G}^A_B=\mathrm{G}^A_B+\frac{1}{2}\beta \l^2
\mathcal{R}^{\,A}_{2\, B}-\frac{1}{8}\beta
\l^2\mathcal{R}_{2}\,\delta^A_B+\Lambda \delta^A_B=0.\label{Ein5d}
\ea
For pedagogical reasons, let us consider the case that the bulk is AdS
with length scale $\l$, so that the metric is
\ba
\d s^2=\d y^2+e^{-2 y/\l} \eta_{\mu \nu}\d x^\mu \d x^\nu,\label{dsback}
\ea
in flat slicing. Then the Einstein equation \eqref{Ein5d} implies the
relation between $\beta$, $\l$ and $\Lambda$.
\ba
\mathcal{G}^A_B
=\frac{1}{\l^2}\(6-3 \beta+\l^2 \Lambda\)\delta^A_B =0.
\ea
The AdS length scale is therefore related to the bulk cosmological
constant
\ba
\Lambda=-\frac{6}{\l^2}+\frac{3\beta}{\l^2}.\label{Lambda}
\ea
In the absence of GB terms, ($\beta=0$), we recover the usual
canonical RS tuning \cite{Randall:1999vf}. In the
presence of GB terms, the bulk can hold an AdS solution {\it without}
the presence of any bulk cosmological constant if $\beta$ takes the
specific value $\beta=2$, as already pointed out in
\cite{Deser:1987uk,Deser:1989jm,Crisostomo:2000bb,Corradini:2004qr}.
There is an AdS
solution in the presence of a {\it positive} bulk cosmological constant when $\beta>2$.
In this paper we explore the possibility of having relatively large GB
terms, {\it ie.} $\beta \gtrsim 1$, but not necessarily $\beta>2$.

The boundary conditions on the brane is given by the analogue of the
Isra\"el matching conditions \cite{Israel:1966rt} in presence of GB terms
\cite{Myers:1987yn,Davis:2002gn}:
\ba
K^\mu_\nu+\frac{\beta \l^2}{3}\(\frac 9 2 J^\mu_\nu- J
\delta^\mu_\nu-3P^\mu_{\ \ \alpha \nu \beta}K^{\alpha
  \beta}+ P^\rho_{\ \ \alpha \rho \beta}K^{\alpha
  \beta}\delta\mupn \)\notag \\
\hspace{-25pt}=-\frac{\kf}{6}\lambda
\delta\mupn-\frac{\kf}{2}\(T\mupn-\frac 1 3 T \delta \mupn\),\hspace{8pt}\label{bdyCond}
\ea
where
\ba
\hspace{-5pt}P^\mu_{\ \ \alpha \nu \beta}&=&
\(R^\mu_\beta q_{\alpha \nu} +
R_{\alpha\nu}\delta^\mu_\beta-R_{\alpha
  \beta}\delta^\mu_\nu-R\mupn q_{\alpha \beta}\)\\
&&+R^\mu_{\ \ \alpha \nu  \beta}-\frac 1 2 R\(\delta^\mu_\beta
q_{\alpha\nu}-q_{\alpha \beta}\delta\mupn\),\notag
\ea
$R^\mu_{\ \ \alpha \nu \beta}$ being the four-dimensional Riemann
tensor induced on the brane. $T\mupn$ represents the four-dimensional
stress-energy tensor associated with the matter fields confined on the
brane $\mathcal{L}_m$.

As for the normal RS model, a brane moving through a pure AdS bulk
will undergo a flat Friedmann-Robertson-Walker (FRW) expansion or
contraction in  the induced geometry. In this case the extrinsic curvature is
$K^i_j=k\delta^i_j$, with
$k=-\frac 1 \l \sqrt{1+\l^2 H^2}$,  $H$ being the Hubble
parameter on the brane. We have as well
$\frac 9 2 J^i_j-J \delta^i_j=- k^3 \delta^i_j$ and
$-3P^i_{\ \ \alpha j \beta}K^{\alpha
  \beta}+ P^\rho_{\ \ \alpha \rho \beta}K^{\alpha
  \beta}\delta^i_j=3H^2 k \delta^i_j$.
For spatially flat cosmologies the boundary condition \eqref{bdyCond} simplifies considerably:
\ba
k\(1+\beta \l^2 \(H^2-\frac 1 3 k^2\)\)=-\frac{\kf}{6}\(\lambda+\rho\),\label{bdyCondBack}
\ea
$\rho$ being the energy density of the matter fields located on the
brane. The modified Friedmann equation on the brane is therefore given
by the solution of:
\ba
\sqrt{1+\l^2 H^2}\left[1+\frac 1 3 \beta \(-1+2 \l^2
H^2\)\right]=\frac{\l \kf}{6} \(\lambda+\rho\)\label{Fried0}.
\ea
In particular, if the brane is empty for the background, $\rho=0$, and
if the brane tension $\lambda$ is fine-tuned to its canonical value:
\ba
\lambda=\frac{6}{\kf \, \l}\(1-\frac 1 3 \beta\),\label{cano}
\ea
the brane geometry becomes Minkowski spacetime ($H=0$) and the brane position
remains static.

\subsection{Linearized Gravity} \label{secLG}
We now consider linear perturbations around the background AdS
geometry (\ref{dsback}). In particular we
will consider the brane (at $y=0$) to be empty for the background and to have a
fine-tuned tension  \eqref{cano}.
We wish to study the metric perturbations sourced by
matter confined on the brane with stress-energy tensor $T\mupn$. In RS
gauge, the perturbed metric is:
\ba
\d s^2=\d y^2+\(e^{-2|y|/\l}\eta\mn+h\mn\)\d x^\mu \d x^\nu\\
\text{with } h^\mu_\mu=0 \text{ and } h\mupn{}_{,\mu}=0,\notag
\ea
where indices are raised with respect to $\eta^{\mu \nu}$.
In this gauge, the Einstein equation in the bulk is:
\ba
\(1-\beta\)\left[e^{2
    y/\l}\Box+\partial_y^2-\frac{4}{\l^2}\right]h\mupn=0, \label{Einpert}
\ea
where $\Box$ is the four-dimensional Laplacian in flat spacetime $\Box=\eta^{\mu\nu}\partial_{\mu\nu}$.
In this gauge, the brane location is no longer fixed. We denote by
$\delta y$ its deviation from $y=0$. We hence work in the
Gaussian Normal (GN) gauge where the brane is ``static":
\ba
\bar{y}&=&y-\delta y(x^\mu),\\
\bar{x}^\mu&=&x^\mu-\(\frac{\l}{2}\(e^{2y/\l}-1\)+\frac{2}{\l \Box}\)\eta^{\mu\nu} \delta y_{,\nu}.
\ea
In this new
gauge, the perturbed metric is:
\ba
\bar{h}\mn(\bar y)&=&h\mn(y)-\l(1-e^{-2y/\l})\delta y_{, \mu\nu}\\
&&-\frac{2}{\l}e^{-2y/\l}\(\eta\mn+\frac 2 \Box \partial\mn\)\delta
y. \notag
\ea
We choose the remaining degrees of freedom in the GN gauge such that on
the brane, the induced metric perturbation $\bar h\mn (0)$ is in de
Donder gauge, hence $R\mupn=-\frac 1 2 \Box \bar{h}\mupn(0)$.
Using this relation, the boundary conditions becomes
\ba
\(1-\beta\)\delta K\mupn+\frac \l 2 \beta\(\Box \bar{h}\mupn-\frac 1
6 \Box \bar h \delta\mupn\)\\=-\frac \kf 2\(T\mupn-\frac 1 3 T
\delta\mupn\),\notag
\ea
where $\delta K\mupn=\frac 1 2 \left[\partial_y +\frac 2
  \l\right]\bar h\mupn$. The boundary condition in RS gauge is
  summarized as:
\ba
\left(1-\beta\right)\left[\partial_y +\frac 2
  \l\right]h\mupn
+\beta \l \Box h\mupn=-\kf\Sigma\mupn,\label{bdy}
\ea
where $\Sigma\mupn$ is a traceless and transverse tensor associated with the
stress-energy tensor $T\mupn$:
\ba
\Sigma\mupn=T\mupn-\frac{1}{3}T\delta\mupn+\frac{1}{3\Box}\eta^{\mu\alpha}T_{,\alpha\nu},\label{Sigma}
\ea
and the perturbation of the brane location is $\delta
y=-\frac{\kf}{6\(1+\beta\)\Box}T$.

Solving the bulk Eq. of motion \eqref{Einpert} with the boundary
condition and the requirement that the perturbations die off at
infinity, one has the solution:
\ba
h\mn(y)=\kf \hat{F} \Sigma\mn,
\ea
with
\ba
\hat{F}=\frac{1}{\sqrt{-\Box}}\frac{K_2\(\sqrt{-\Box}\l
e^{y/\l}\)}{\(1-\beta\)K_1\(\sqrt{-\Box}\l\)+\beta \sqrt{-\Box} \l
K_2\(\sqrt{-\Box}\l\)},\hspace{5pt}
\ea
where $K_n$ is the Bessel function.
We may decompose this expression into the zero mode and the infinite tower of
KK corrections for the induced metric perturbation on the brane:
\ba
&\bar h\mn=-\frac{2 \kf}{\l(1+\beta)\Box}\(T\mn-\frac 1 2 T
\eta\mn\) \notag \hspace{90pt}\\
&+\frac{1-\beta}{1+\beta}\frac{K_0\(\sqrt{-\Box}
\l\)}{(1+\beta)K_1\(\sqrt{-\Box}\l\)+\beta \sqrt{-\Box} \l
K_0\(\sqrt{-\Box}\l\)} \frac{\kf}{\sqrt{-\Box}} \Sigma\mn. \label{KKcorr}
\ea
It is therefore clear from this result, that when $\beta<1$, the KK
tower gives a positive correction to the zero mode which makes
gravity stronger at short distances, whereas when $\beta>1$, the KK
corrections will make gravity weaker at short distances. When $\beta=1$, linearized gravity appears purely
four-dimensional on the brane ({\it ie}. without any KK correction) \cite{Crisostomo:2000bb}.

To make this argument more precise, we consider a point-like source
of mass $M$ on the brane: $T_{00}=M \delta(r)$, $T_{ij}=0$. This local
source generates the following metric perturbations:
\ba
\bar h_{00}&=&\kf M\(\frac 1 2 V_0+\frac 2 3 V_{KK}\)\\
\bar h_{ij}&=&\kf M\(\frac 1 2 V_0+\frac 1 3 V_{KK}\)\delta_{ij}
\ea
where the contribution $V_0$ from the zero mode is
\ba
V_0
=\int_0^\infty \d k\frac{k \sin kr}{2 \pi^2r}\frac{2}{ \l
(1+\beta)k^2}=\frac{1}{2 \l (1+\beta)\pi
r},
\ea
and the contribution from the KK modes is (Cf. Appendix \ref{app} and Eq. \eqref{VKK})
\ba
V_{KK}=\frac{1-\beta}{\(1+\beta\)^2}\frac{\l}{4\pi r^3}+\mathcal{O}\(\l^3/r^5\).
\ea
The total gravitational potential  generated by a source point of
mass $M$ on the brane is therefore of the form:
\ba
V(r)&=&-\frac{\kf M}{2}\(V_0+V_{KK}\)\notag \\
&=&-\frac{\kf M}{4 \pi \l (1+\beta)
r}\(1+\frac{1}{2}\frac{1-\beta}{1+\beta}\frac{\l^2}{r^2}+\cdots\).
\label{V(r)}
\ea
This result, already obtained in Ref. \cite{Neupane:2001st} is completely
valid even if $\beta>1$. When $\beta=1$, we recover the previous
result that $V_{KK}=0$, but for $\beta>1$, the KK terms give a
negative contribution to the zero mode, and gravity hence appears
weaker at short distances. This can be seen in the numerical
results from Ref. \cite{Deruelle:2003tz}. One can evaluate the expression \eqref{VKK1}
for $V_{KK}$ numerically to understand how the gravitational potential behaves as a
function of $\beta$,
\ba
\hspace{-8pt}V(r)
=-\frac{\kf M}{4 \pi \l (1+\beta)
r}\(
1+\int_0^{\infty} \hspace{-3pt} \d m\ e^{-m r} \rho_\beta (m\l)\),\label{V(r)2}
\ea
where we write
\ba
\rho_\beta(x)=\frac{1+\beta}{1-\beta} \frac{2}{x \pi^2}\ \frac{1}{\tilde J_\beta(x)^2+\tilde
Y_\beta(x)^2},
\ea
using the same notation as in appendix \ref{app}.
The integral in \eqref{V(r)2} is always negative for $\beta>1$ but
remains greater than $-1$, for any value of $\beta$ and any distance
$r$. It is therefore clear
 that gravity is weaker for $\beta>1$ but remains attractive.

\begin{figure}
 \centering
 \includegraphics[width=.9\columnwidth]{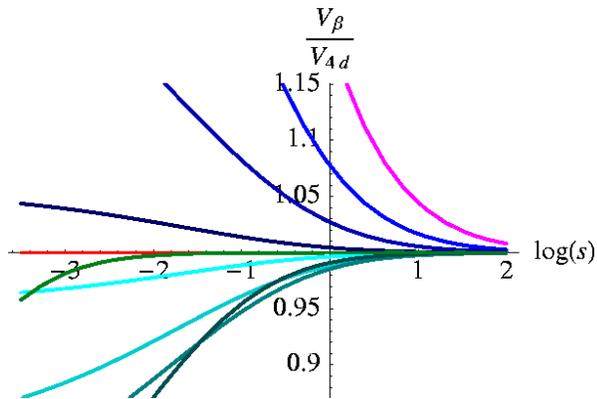}
 \caption{Ratio of the gravitational potential to the 4d one $V_\beta/V_{4d}$ as a
 function of $\log(s)=\log(r/L)$.
The pink curve (top one) represents the pure RS case ($\beta=0$),
and the blue curves (the ones below) are for $\beta=0.3,  0.6,
0.9$. The horizontal red curve represents the purely 4d case
$\beta=1$, and the green curves below (from lightest to darkest) are
for $\beta=1.1, 1.5, 2, 100$. The curve with $\beta=100$ differs from
 the 4d only at very short distances $s\sim 10^{-3}$.}
 \label{pic1}
\end{figure}

Writing
$L=\l(1+\beta)$, the four-dimensional gravitational coupling constant
can be expressed in terms of the fifth one as $\kappa_4^2=\kf/L$. $L$
is usually assumed to be of order of $0.1$mm or smaller. We indeed expect to observe
a deviation from four-dimensional gravity at distances of order of $L$, but
the E\"ot-Wash Short-range experiment has measured the strength
 of gravity for distances slightly smaller than $0.2$mm and have not observed any
 deviation from the Newton's law. Recent experiments are still
 testing gravity at sub-millimeter scales \cite{Hoyle:2000cv}.
In terms of this fundamental scale, the gravitational
potential generated by a source of mass $M$ is
\ba
\hspace{-2pt}V(s)=-\frac{\kappa_4^2M}{2 \pi L s}
\(1+
\int_0^{\infty} \!\! \d x \, e^{-x (1+\beta)s}\rho_\beta(x)\),\label{Vnum}
\ea
where $s$ is the distance to the source measured in units of $L$:
$s=r/L$.
The integral in \eqref{Vnum}
vanishes in the large GB term limit, $\beta\rightarrow
\infty$, (if we take the limit $\beta\rightarrow \infty$, {\it before}
taking the limit $s\rightarrow 0$),
leading to purely four-dimensional gravity in that
limit. In order to understand the effect of the KK corrections, we
integrate the integral numerically.
Fig. \ref{pic1} represents the ratio of the gravitational potential to
the four-dimensional one for different
values of $\beta$. Gravity does indeed appear stronger than usual four-dimensional
gravity for $\beta<1$, and weaker if $\beta>1$.

%
%
%
%
%
\begin{figure}
 \centering
 \includegraphics[width=.9\columnwidth]{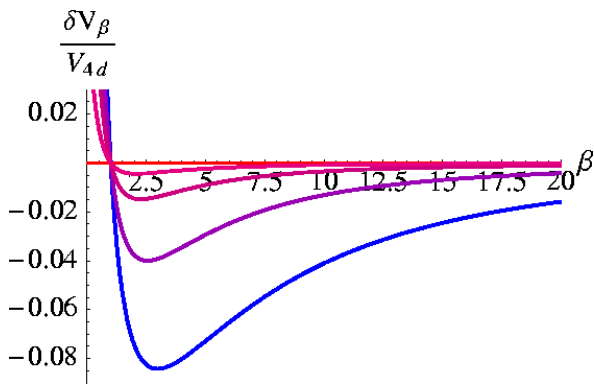}
 \caption{Relative departure of the gravitational potential to the 4d
 one $\( V_\beta-V_{4d}\)/V_{4d}$ as a
 function of the parameter $\beta$, for different values of $s$.
 The red horizontal curve is the purely 4d case and the curves below from red to blue (or from top to
 bottom) are evaluated at $s=0.25, 0.5, 1, 2$.}
 \label{pic3}
\end{figure}
%
As can be seen from Fig. \ref{pic1}, gravity can only
be mildly weakened by the presence of GB terms with $\beta>1$, and in
particular there is an upper bound on the modification of gravity, as
is more clear in Fig. \ref{pic3} where the ratio of the
correction to the four-dimensional potential is represented as a function of $\beta$, for
different distances $s$ to the point source.

Near the source, relative to the fundamental scale $L$,
the integral in \eqref{Vnum} goes as
\ba
\int_{0}^{\infty} \d x  \rho_\beta(x)=-\frac{1-\beta}{2\beta},
\ea
for $\beta>0$, so that very close to the source, the gravitational potential is
\ba
V(r)\xrightarrow{r\rightarrow0} -\frac{\kappa_4^2M}{2 \pi
  r}\frac{1}{2 \beta}. \label{Vasym}
\ea
Near the source, the gravitational potential therefore
becomes four-dimensional again, but with a coefficient which is now
changed.  We recover the usual coefficient when $\beta=1$, and we see
that as $\beta \rightarrow 0$, {\it ie.} in the
RS limit, the coefficient diverges and the behaviour would be
completely different from the case $\beta \neq 0$.

\subsection{Implications for Cosmology}
\label{secIC}
We now study the implications of a GB braneworld with $\beta \ge 1$ for cosmology.
Previous works have studied the consequences of the GB term
in cosmology \cite{Nojiri:2000gv,Dufaux:2004qs,Tsujikawa:2004dm}, but most of these studies have focused on the regime
where $\beta\ll1$, as one would expect if this term arose from
quantum/string corrections.

From the boundary condition \eqref{bdyCondBack} in the background,
the modified Friedmann equation is:
\ba
H^2=\left\{
\begin{array}{ll}
\frac{1}{\l^2 \beta}\left[ |1-\beta| \cosh \frac 2 3 \chi -1
 \right]& \text{if}\ \beta\ne 1\\
 \frac{1}{\l^2}\left[\(1+\frac{\rho}{\lambda}\)^{2/3}-1\right] & \text{if}\ \beta=
 1,
 \end{array}\right.\label{Fried}
\ea
with
\ba
\sinh \chi =\frac{\l \kf
\lambda}{2}\sqrt{\frac{\beta}{2(1-\beta)^3}}
\(1+\frac{\rho}{\lambda}\)\ \ \text{for } \beta<1\\
\cosh \chi =\frac{\l \kf
\lambda}{2}\sqrt{\frac{\beta}{2(\beta-1)^3}}
\(1+\frac{\rho}{\lambda}\)\ \ \text{for } \beta>1,
\ea
which generalizes the result in
\cite{Dufaux:2004qs,Tsujikawa:2004dm} to more general cases
of $\beta$.

\subsubsection{Low-energy regime}
\label{sec lowenergy GB}

At low energies compared with the brane tension, $\rho\ll \lambda$,
the Friedmann equation \eqref{Fried} recovers a four-dimensional
behavior: $H^2\simeq\frac{\kappa_4^2}{3}\rho$, and the
higher-dimensional nature of the theory won't affect the
cosmological evolution within this range of energy.
This approximation is valid as long as $\rho\ll \rho_\lambda$, where $\rho_\lambda=\frac{2(3-\beta)}{\l
\kf}=2(3-\beta)(1+\beta)\(L \kappa_4\)^{-2}$. Assuming that $L=\(1+\beta\) \l$ is of the order of $0.1$mm, the critical density is then of order $\rho_\lambda \simeq 2(3-\beta)(1+\beta)
M^4_{\text{EW}}$, where $M_{\text{EW}}$
is the electroweak scale. For $\beta \sim 1$, this takes an order of
magnitude of $\rho_\lambda\simeq 8 M^4_{\text{EW}}$.

However one might be worried here that this low-energy effective theory
is never a good approximation when
the parameter $\beta$ is such that $\lambda=0$ (ie. when $\beta \simeq
3$). This is however never the case. Expanding the expression \eqref{Fried} for small $\rho$, we
get
\ba
H^2=\frac{\kappa_5^2}{3}\rho+\frac{1}{36}\frac{1-3
  \beta}{\(1+\beta\)^3}L^2 \kappa_4^4 \rho^2+\mathcal{O}\(L^4 \kappa_4^6 \rho^3\),
\ea
so that the low-energy effective-theory is actually a good
approximation as long as $\rho\ll \rho_c$, where
$\rho_c=\frac{12\(1+\beta\)^3}{ \(3\beta-1\)L^2\kappa_4^2}\sim
\frac{12\(1+\beta\)^3}{\(3\beta-1\)}M^4_{\text{EW}}$. Such that when
$\beta=1$, the real bound is $\rho_c\sim 48 M^4_{\text{EW}}$ and when
the brane tension vanishes, $\beta\simeq 3$, $\rho_c\sim 96
M^4_{\text{EW}}$.
The low-energy effective limit is therefore obtained
even before what would be expected from the requirement that $\rho\ll
\lambda$, if $\beta>1$.

In order to recover four-dimensional behavior right after inflation, we
need the reheating temperature to be much smaller than this critical
scale. Typical inflationary models occur at energies much larger
than this scale, but there exist models which occur at energy scales
of order of the electroweak scale, such as Ref.
\cite{vanTent:2004rc}. In this specific model, the reheating
temperature is of order $T_{\text{Re}}\sim 0.43
M_{\text{EW}}$, so the energy scale at the end of inflation is
$\rho_{\text{Re}}\sim 0.03 M^4_{\text{EW}}$.
At the end of this inflationary model, the
energy scale will be of order $\rho_{\text{Re}} \sim 7.
10^{-4}\rho_c$, if $\beta\sim 1$.
However, it might still be possible to imagine reheating mechanisms
that could accomodate this modification of the standard Friedmann
equation. The only real constraint is therefore that nucleosynthesis
begins at energies below this critical scale. The energy scale at the
beginning of nucleosynthesis is usually taken to be of order
$\rho_{n}\sim 10^{-28}M^4_{\text{EW}}$, which is clearly much below
the critical value.

The low-energy regime will therefore be a good
approximation and the presence of the extra-dimension or the
GB terms won't affect cosmology at scales after inflation.
We therefore argue that having large GB terms in the bulk
is not incompatible with observations.

Furthermore, it has been shown that the presence of these GB terms might
actually help to resurrect steep inflation driven by an exponential potential, as pointed out in
\cite{Tsujikawa:2004dm}, as well as quadratic and quartic potential
inflationary models. The presence of $\rho^2$ terms in the Friedmann
equation of the RS model, have raised the possibility of having a
steep inflation scenario on the brane \cite{Maartens:1999hf}. In the present case, we find at high energies
\be
H^2 \approx \frac{1}{\l^2} \left( \frac{\l\kf}{4\beta}\rho
\right)^{2/3}
\ll \frac{\kf}{4\beta \l} \rho.
\ee
It is therefore less steep than normal inflation. The model of steep
inflation will hence be invalid in this case.

\subsubsection{Gravitational waves}

To finish this section on cosmological consequences, we comment on the
behaviour of gravitational waves during inflation. It has been shown
in Ref. \cite{Dufaux:2004qs}, that the presence of GB terms lead to a
modified amplitude for the tensor perturbations. In particular, their
amplitude is shown to first increase with energy scale, just as in the
RS scenario, but as second stage, to decrease asymptotically.
Using the same notation as in  \cite{Dufaux:2004qs}, the tensor
amplitude is indeed shown to be of the form
\ba
A_T^2&=&\kappa_4^2\(\frac{H^2}{2\pi}\)^2 F_\beta^2\(\l H\)\\
F_\beta^{-2}(x)&=&\sqrt{1+x^2}-\(\frac{1-\beta}{1+\beta}\)x^2
\sinh^{-1}\frac 1 x.
\ea
In the usual four-dimensional theory, the second term in the
expression for $F_\beta^{-2}$ vanishes, and $F^{2}$ decreases
monotonically as a function of the energy. In the RS case, however,
$F_0^2$ increases monotonically. This behaviour is perturbed by the
presence of GB terms, which usually follow the same behaviour as for
RS at low-energy, before starting to decrease as in the
four-dimensional case. The gravitational waves therefore increase
above standard level before decreasing asymptotically as mentioned in
\cite{Dufaux:2004qs}. However, if the GB are important enough, in
particular if $\beta>1$, they won't have time to follow the RS
behaviour before adopting the four-dimensional one, and the usual
behaviour pointed out in \cite{Dufaux:2004qs} will no longer be valid.
The ratio of tensor to scalar perturbations amplitude will therefore
remain close to the four-dimensional one within this regime.

\subsection{Stability}
When $-1<\beta<1$, no ghosts are present in the theory, see Ref.
\cite{Cho:2001su}.
It has been shown in \cite{Deser:1987uk,Maeda:2003vq}, that in
absence of any cosmological constant, EH-GB
theories in more than four dimensions admit an AdS solution, beside the
flat-Minkowski solution. This is indeed precisely the behaviour we get in \eqref{Lambda} when
$\beta=2$. Although this solution seems a priori well-defined and
follows from a consistent theory of gravity, it has been shown in
Ref. \cite{Deser:1989jm}, that it corresponds to a gravitationally
unstable solution. The ADM mass of any massive object living in the
five-dimensional bulk will be negative which signals an instability \cite{Gleiser:2006yz,Cai:2001dz},
although interesting work in Ref. \cite{Gibbons:2004au} has suggested that even a negative ADM mass object could be stable with suitable
boundary conditions.
Writing $\alpha=\beta\, \l^2$, the equation relating the AdS length
scale to the bulk cosmological constant is then quadratic in $\l^2$:
$\Lambda=-6/\l^2+3\alpha/\l^4$ such that there are two kind
of solutions for $\l^2$:
\ba
\l^2=-\frac{3}{\Lambda}\(1\pm\sqrt{1+\frac{\alpha \Lambda}{3}}\),
\ea
the solution with the upper sign corresponds to the solution
for which we recover the RS limit when $\alpha=0$.
This represents the gravitationally stable branch. In order to
recover the self-interacting AdS solution in absence of the
cosmological constant, one should however consider the solution with
the lower minus sign such that $\l^2=\alpha/2$.

Any solution with $\beta>2$, {\it ie.} with $\Lambda>0$, must come
from the solution with the lower sign, {\it ie.} the unstable solution.
For $\Lambda<0$ ({\it ie.} for $\beta<2$), the maximum value $\alpha$
can take in any of the two branches is
$\alpha<\alpha_c=-3/\Lambda$. This translates into a maximal value for
the parameter $\beta$: $\beta<\beta_c=\(1\pm\sqrt{1+\alpha
\Lambda/3}\)^{-1}$. Any solution with $\beta>1$ must therefore have
originated from the branch with the lower sign (unless $\alpha<0$)
which appears gravitationally unstable when
positive matter is introduced in the bulk, as mentioned in \cite{Deser:1989jm}.
This can easily be understood from the Einstein
equation \eqref{Einpert}. If some matter with stress-energy
$\tau^{b}\mn$ was introduced in the bulk,
the right hand side of the Einstein equation would be of the form
$\(1-\beta\){} ^{(5)\!}\Box h\mn=-2 \kf \tau^{b}\mn$.
When $\beta>1$, matter will couple to gravity with the wrong sign,
leading to naked singularities. When $\beta>1$, one should therefore
consider matter to be introduced with the opposite sign in the bulk, for the theory to make
sense gravitationally. We may note that the back-reaction from bulk
gravitational waves won't produce any instability, since they will
enter with the correct sign in the modified Einstein equation.

 In what follows, we take a slightly different approach, and present an
alternative model where the bulk action remains the conventional
EH one, with the conventional sign for gravity in the bulk. Instead, the
price to pay will be to invert the contribution from the brane
action. Despite this, conventional gravity is recovered on the brane at low energies.

\section{Alternative approach} \label{sec alt}
\subsection{EH-RS model}
We consider in what follows the alternative approach, where the bulk
gravitational action is the same as in the RS model, but the
boundary action is unconventional:
\ba
S^{(\text{5d})}&=&\frac{1}{2 \kf}\int \d ^5x\sqrt{-g} \Bigl[
\mathrm{R}-2{\Lambda} \Bigr]-S^{(\text{4d})}_b\label{newmodel}\\
S^{(\text{4d})}_b&=&\int_{\text{brane}}\hspace{-10pt}\d^4 x \sqrt{-q}\left[\frac{\tilde{\beta}}{2\kt}
R+\mathcal{L}_m+{\tilde{\lambda}}-2K\right],\notag
\ea
where we take $\kt=\kf/\l$, and we consider $\tilde{\beta}>0$. We consider
the brane tension to be fine-tuned to the canonical value ${\tilde{\lambda}}=6/\kf
\l$.
This can be seen as a combination of the RS model and the Dvali-Gabadadze-Porrati (DGP) one
\cite{Dvali:2000rv} (although we do not concentrate on the
self-accelerating branch of the DGP model), with the ``wrong" sign
for the brane action. For a realization of this model from string
theory, see Ref. \cite{Antoniadis:2002tr}.

We can analyze this model in the same manner as the RS model by
replacing the usual four-dimensional stress-energy tensor for matter
field by:
\ba
T^{\text{RS}}\mn\rightarrow -T\mn+\frac{\tilde{\beta}}{\kt}G\mn ,
\ea
where now $T\mn$ is the stress-energy of matter field on the brane:
$T\mn=-\frac{2}{\sqrt{-q}}\frac{\delta}{\delta
q^{\mu\nu}}\sqrt{-q}\mathcal{L}_m $.

The low-energy effective theory in the RS model is
$G\mn=\kt T^{\text{RS}}\mn$,
and therefore gets replaced by
\ba
G\mn=\frac{\kt}{\tilde{\beta}-1}T\mn\label{Geff lowenergy}
\ea
in the new theory \eqref{newmodel}. The parameter $\tilde{\beta}$ should
therefore satisfy $\tilde{\beta}>1$, for matter to couple to gravity the right way.
In the IR regime, this theory will therefore be completely
consistent.

Considering a homogeneous and isotropic background, the Isra\"el
matching condition in the RS case is $\rho^{\text{RS}}={\tilde{\lambda}}\(\sqrt{1+\l^2
H^2}-1\)$. Using the transformation $\rho^{\text{RS}}\rightarrow -\rho+{\tilde{\lambda}} \l^2 H^2/2$,
this equation becomes quadratic for $H^2$, and as pointed out in the
DGP model, there are therefore two branches for the solution:
\ba
H^2=\frac{2(\tilde{\beta}-1)}{\tilde{\beta}^2\l^2}\left[
-1+\frac{\tilde{\beta}}{\tilde{\beta}-1}\frac{\rho}{{\tilde{\lambda}}} \pm \sqrt{1+2
\frac{\tilde{\beta}}{(\tilde{\beta}-1)^2} \frac{\rho}{{\tilde{\lambda}}}}\right]. \label{FRW2}
\ea
The branch with the upper sign is the one that recovers the usual four-dimensional Friedmann equation at
low-energy, $\rho\ll {\tilde{\lambda}}$. The other solution, presents a
self-accelerating behaviour at low-energy $\l^2 H^2\simeq
4(1-\tilde{\beta})/\tilde{\beta}^2$, and has been pointed out in
\cite{Dvali:2000rv}, as a potential explanation for the cosmological
constant problem (note that this solution could only be possible here
if $\tilde{\beta}<1$, which is not the case we consider here).
Recent arguments confirm the instability of this branch, see
Ref. \cite{Koyama:2005tx,Charmousis:2006pn} for recent reviews on the subject.

\subsection{Linearized Gravity}

In what follows, we concentrate on the solution \eqref{FRW2} with the upper sign, and study
linearized gravity around the five-dimensional AdS, with
$\rho=0$. We follow the same procedure as in section \ref{secLG},
using the same notation.
We study the perturbations generated by the presence of some matter
source $T\mn$ on the brane. In the RS model, the presence of this
source generates the induced perturbations on the brane:
\ba
\bar{h}\mn^{\text{RS}}\hspace{-5pt}&=&\hspace{-5pt}\frac{\kf}{\sqrt{-\Box}}\frac{K_2\(
  \sqrt{-\Box} \l\)}{K_1\(
\sqrt{-\Box}\l\)}\(T\mn^{\text{RS}}-\frac 1 3
T^{\text{RS}}\eta\mn+\frac{1}{3\Box}T^{\text{RS}}_{,\mu\nu}\)\notag\\
\hspace{-5pt}&&\hspace{-5pt}-\frac{\kt}{3\Box}\(T^{\text{RS}}\eta\mn+\frac{2}{\Box}T^{\text{RS}}_{,\mu\nu}
\),
\ea
in de Donder gauge \cite{Garriga:1999yh,deRham:2004yt}.
Using this result, and performing the transformation
\ba
T^{\text{RS}}\mn&\rightarrow&-T\mn+\frac{\tilde{\beta}}{\kt}\(R\mn-\frac 1 2 R
\eta\mn\)\notag\\
&=&-T\mn+\frac{\tilde{\beta}}{2(\tilde{\beta}-1)}T\eta\mn-\frac{\tilde{\beta}}{2\kt}\Box
\bar{h}\mn,
\ea
we obtained the induced perturbations on the brane
\ba
\bar{h}\mn=-\frac{2\kt}{(\tilde{\beta}-1)}\Biggl[\frac{1}{\Box}\(T\mn-\frac 1 2 T \,
\eta\mn\) +\frac{\l
\hat{K}_{\tilde{\beta}}}{\sqrt{-\Box}}\Sigma\mn\Biggr]\label{h newmodel}
\ea
where
\ba
\hat{K}_{\tilde{\beta}}=\frac{K_0\(\sqrt{-\Box}\l\)}{2(\tilde{\beta}-1)K_1\(\sqrt{-\Box}\l\)+\tilde{\beta}
\sqrt{-\Box}\l K_0\(\sqrt{-\Box}\l\)},
\ea
and $\Sigma\mn$ as defined in \eqref{Sigma}.

At low energies, we therefore recover the four-dimensional behaviour
as expected, with $\kappa_4^2=\kt/(\tilde{\beta}-1)$. This theory only
makes sense if $\tilde{\beta}>1$, in which case the KK corrections in
\eqref{h newmodel}, come with a negative sign, similarly as in
\eqref{KKcorr}, giving rise to a weaker gravity at
small scales.

This theory will therefore behave similarly at
small scales as the GB-RS theory described previously, in the regime
where $\tilde{\beta}>1$.  Far away from the brane,
this theory will however appear well-defined, and will not present
the instabilities pointed out before.

\subsection{Brane energy conditions}

In this section we demonstrate that the brane satisfies, in certain
limits, the usual energy conditions. We define the total energy of the
brane to be the sum of the canonical brane tension ${\tilde{\lambda}}$ and contributions from the EH
term and brane matter:
\ba
T^{\text{tot}}\mn&=&\frac{2}{\sqrt{-q}}\frac{\delta \sqrt{-q}}{\delta
  q^{\mu\nu}}\(
\frac{\tilde{\beta}}{2\kt}R+\mathcal{L}_m+{\tilde{\lambda}}\)\\
&=&\frac{\tilde{\beta}}{\kt}G\mn-T\mn- {\tilde{\lambda}}\, q\mn.
\ea
First of all, at low energies the
total brane stress energy is equal to the brane tension plus a
term proportional to the stress energy of matter. Using the relation
\eqref{Geff lowenergy}, valid at low energies, we have
\be
T^{\text{tot}}\mn=-{\tilde{\lambda}} \, q\mn+\frac{1}{\tilde{\beta}-1}T\mn.
\ee
If the brane matter $T_{\mu\nu}$ satisfied either the weak or null
energy conditions, then it is immediately clear that
$T^{\text{tot}}_{\mu\nu}$ satisfies the same conditions.
We can also see what happens at high energies, but at long wavelengths
using the separate universes approach \cite{Guth:1982ec,Starobinsky:1986fx} whereby long wavelength
perturbations are modeled locally as a FRW universe with
curvature. Then the null energy condition amounts to $p+\rho \ge 0$
and the weak energy condition has the additional constraint $\rho \ge
0$. Using the effective Friedmann equation for the brane (\ref{FRW2})
we can infer the total energy density on the brane
\ba
\rho_{\text{tot}}&=&\frac{3\tilde{\beta}}{\kt}H^2-\rho+{\tilde{\lambda}}\notag\\
&=&\frac{{\tilde{\lambda}}}{\tilde{\beta}}+{\tilde{\lambda}} \frac{(\tilde{\beta}-1)}{\tilde{\beta}}
\sqrt{1+2\frac{\tilde{\beta}}{(\tilde{\beta}-1)^2} \frac{\rho}{{\tilde{\lambda}}}}\ .
\ea
It is clear that $\rho_{\text{tot}}$ satisfies $\rho_{\text{tot}} \ge {\tilde{\lambda}}/{\tilde{\beta}}$
and so provided $\rho$ satisfies the null energy condition then
$\rho_{\text{tot}}$ satisfies the weak and null energy conditions.

We can extend this argument by considering the effect of taking the
bulk to be Schwarzschild-AdS with either a positive or negative ADM
mass. In this case we recover a similar result as long as the brane
is sufficiently far away from the bulk black hole ({\it ie.} outside the
Schwarzschild radius of this object if its ADM mass is positive).

\subsection{Implications for Cosmology}
As mentioned previously, provided we consider the stable branch of
the solution \eqref{FRW2}, the usual four-dimensional Friedmann
equation is recovered at low-energy $H^2=\kappa_4^2/3\, \rho$, when
$\rho\ll 2 (1-\tilde{\beta})^2 {\tilde{\lambda}}$, where the four-dimensional gravitational coupling
constant is as defined previously
$\kappa_4^2=\kt/(\tilde{\beta}-1)$.
This low-energy will therefore be satisfied as long as
$\rho\ll 12 (\tilde{\beta}-1)\(\l \kappa_4\)^{-2}$, so that if
$(\tilde{\beta}-1)\sim 1$, this constraint will be satisfied for
$\rho\ll 12 M_{\text{EW}}^4$. Similarly as in section \ref{sec lowenergy
  GB}, this low-energy regime, will be consistent with scenarios of
inflation with a very small reheating temperature, such as the one
proposed in Ref. \cite{vanTent:2004rc}.
However, this model also recovers a
four-dimensional behaviour at high-energy $H^2\simeq 2
\kt/{\tilde{\beta}} \, \rho$, when $\rho\gg {\tilde{\lambda}}\sim M_{\text{EW}}^4$. Within this limit,
cosmology will not be perturbed, and the expression for the slow-roll
parameters will remain very similar to the usual four-dimensional ones
\ba
\epsilon =-\frac{\dot H}{H^2}\simeq
\frac{\tilde{\beta}}{2\kt}\frac{V'{}^2}{V^2},\hspace{20pt}
\eta=\frac{V''}{3H^2}\simeq\frac{\tilde{\beta}}{\kt}\frac{V''}{V}.
\ea
In the limit $\tilde{\beta}\rightarrow 0$, the relations are no longer valid
since the Friedmann equation becomes linear at high-energy $H^2\sim
 \rho^2/{\tilde{\lambda}}^2$ in the pure RS model.
We therefore have a similar conclusion that the
scenario of steep inflation as is usually possible in the pure RS
model will no longer give a consistent scenario for inflation in
this model.

\section{Stability of compact stars}
\label{secOI}
The astrophysical implications of the modification of gravity at small
scales in typical braneworld models have been studied in particular in
\cite{Germani:2001du,Wiseman:2001xt}. Although the vacuum solution
around a massive object confined in four-dimensions is no longer the
Schwarzschild metric \cite{Germani:2001du},
neutron stars do not seem to be affected by the higher order
corrections in the brane according to the numerical studies of
\cite{Wiseman:2001xt}. In what follows, we give some brief comments on
how the stability
of white dwarf and neutron stars can be improved by the presence of GB
terms in the bulk, or in the alternative scenario presented in the previous section.

Following the many-body analysis of Ref. \cite{Azam:2005dw}, when
gravity on the brane is modified, the Chandrasekhar
{\it ground} state (see Ref. \cite{Chandrasekhar:1984}) does not seem
to be affected by the RS corrections.
However, a more precise analysis performed very recently,
shows the existence of another unbounded minimum of the energy
functional at very small radius, when RS corrections are present \cite{Azam:2006pk}.
The existence of this unbounded minimum makes the Chandrasekhar
minimum metastable, although the tunneling probability to the
new ground state is exponentially suppressed (see
Ref. \cite{Azam:2006pk} for details of the argument and the
computation), and it is not clear whether the argument makes sense
within the Schwazschild radius of the star.

In what follows, we review the argument when the KK modes bring a
negative correction to the zero mode and gravity becomes weaker at
short distances.
For that we follow closely the analysis of \cite{Azam:2006pk}
and use the same notation. In particular $M_0$ and $R_0$ are the
typical mass and radius of helium white dwarfs. We consider a star
with mass $M=\bar{M}M_0$ and radius $r=\bar{r}R_0$. We write
$x_F=\bar{M}^{1/3}/\bar{r}$.

In the pure RS scenario, the first order correction
to the gravitational potential, leads  to a short range correction of the energy
functional $\Xi_c $, going like $\l^2/r^{3}$. We write this correction
$\gamma_0 x_F^3$, with $\gamma_0\sim \l^2/R_0^2$, the exact coefficient in
the expression of $\gamma_0$ depends
on the star parameters, but is of order $1$, and is always positive in
the RS model, since the KK corrections give positive correction to the
zero mode and make gravity
stronger at short-distances. The modified energy functional is then of
the form
\ba
\Xi_c=\frac{3 \pi^2
  \bar{\mathcal{E}}(x_F)}{x_F^3}-1-\bar{M}^{2/3}x_f-\gamma_0 x_F^3,\label{Xi1}
\ea
where the two first terms are related to the kinetic energy of the
free Fermi electron gas, and the function $\bar{\mathcal{E}}(x_F)$ is
such that, for very small radius, $x_F>>1$, $\frac{3 \pi^2
  \bar{\mathcal{E}}(x_F)}{x_F^3}\rightarrow \frac{3}{4} x_F$.
The third term represents the contribution from the Newton
gravitational potential and the last term the leading correction from
the KK modes in the RS scenario.

Below some critical mass, this energy
functional has a unique minimum which is the Chandrasekhar ground
state, if the mass $\bar{M}$ is below some critical value.
However, for very small radius, the energy functional is
\ba
\Xi_c\xrightarrow{x_F\gg1}\(\frac{3}{4}-\bar{M}^{2/3}\)x_F-1-\gamma_0 x_F^3,
\ea
such that in presence of RS corrections, ($\gamma_0>0$), the functional
goes to infinite negative values as the radius goes to zero, the
Chandrasekhar vacuum is then no longer the ground state, and is
metastable. This is a simple consequence of the fact that gravity
becomes stronger at short distances in that model, and the kinetic
energy is no longer able to compensate the increased gravitational
potential. However the situation is different if gravity is weaker at
short distances, and as one would expect, the stability of the star is
then increased in such a situation.

We now examine the same situation when GB terms are included in the
bulk (the situation in the alternative scenario \ref{sec alt}, is
completely analogous). In both scenarios, when $\beta>1$ (or $\tilde{\beta}>1$), gravity is
indeed weaker at short distances and we
don't expect the previous situation to occur.
The sign of the leading correction to the
gravitational potential is indeed negative as we have shown in
\eqref{V(r)}. The leading correction in
\eqref{Xi1}, will therefore have the opposite sign since the
corrections will now be modified to:
$\gamma_\beta \, x_F^3=\(1-\beta\)/\(1+\beta\)\gamma_0 x_F^3\sim\(1-\beta\)/\(1+\beta\)\l^2/r^3 $,
such that when $\beta>1$, $\gamma_\beta$ is now negative.
In that case, the argument presented above will no longer be valid. To be more
precise, we explore the same situation in the limit $x_F\gg1$ when GB
terms are present. As pointed out in \eqref{Vasym}, the modification
to the gravitational potential will recover a $r^{-1}$ behaviour, such
that the corrections will now be of the form $\gamma_\beta x_F$.
The energy functional will therefore be
\ba
\Xi_c\xrightarrow{x_F\gg1}\(\frac{3}{4}+\frac{\beta-1}{\beta+1}\gamma_0-\bar{M}^{2/3}\)x_F-1.
\label{Xibeta}
\ea
When the contribution from the GB
terms is below a critical value $\beta<\beta_c<1$, where $\beta_c$ is
such that the term in bracket in \eqref{Xibeta} vanishes,
 the situation is similar to the one pointed out in
\cite{Azam:2006pk}, and the  Chandrasekhar vacuum is therefore
metastable. However, when $\beta$ is important, $\beta>\beta_c>1$,
there is no longer any unbounded minimum and the Chandrasekhar vacuum
remains the ground state of the theory. The presence of important GB
terms can therefore help recovering a four-dimensional behaviour which
can be broken in a pure RS scenario. The same argument will be valid
when the alternative approach of section \ref{sec alt} is instead
considered.

\section{Summary}
\label{secsum}
The observation of weaker gravity at short scales could present a real
challenge to theoretical physics.
Having a gravitational potential which falls off slower than
$1/r$ is usually hard to obtain without the presence of ghosts in the
theory, or considering the exchange of massive spin-1 bosons which
are highly constrained by experiments on the Equivalence Principle.
Braneworld models usually lead to a gravitational potential
which is stronger at short distances, due to the additional
contribution from the KK modes. In this work we presented two possible
models where the KK modes contribute with an opposite sign to the
usual four-dimensional potential, leading to weaker gravity at short
distances. This comes at the price of introducing instabilities in the
theory which may no longer be quantized the same way as ordinary
four-dimensional gravity at high-energy. However, at low-energy on the
brane, we recover a four-dimensional behaviour and the
theory remains well-defined in that regime. As a test of the proposed
models, we studied the implications for cosmology and showed that
no distinctions from four-dimensional cosmology will be observed after the beginning of
nucleosynthesis.
Within these models, the stability of white dwarfs and neutron stars
against gravitational collapse is typically improved since the gravitational potential is weaker near
the center of the star.


\section*{Acknowledgements}
We wish to thank L.~Boyle and N.~Shuhmaher for interesting conversations.
CdR is funded by a grant from the Swiss National Science Foundation.
The work of TS was supported by Grant-in-Aid for Scientific
Research from Ministry of Education, Science, Sports and Culture of
Japan(No.13135208, No.14102004, No. 17740136 and No. 17340075) and
the Japan-U.K. Research Cooperative Program.  AJT is supported in part by
US Department of Energy Grant DE-FG02-91ER40671.

\appendix
\begin{widetext}
\section{Contribution from the KK modes}
\label{app}

Using the result from \eqref{KKcorr}, the contribution of the KK
modes to the gravitational potential is
\ba
V_{KK}=\frac{1-\beta}{1+\beta}\int \d k \frac{k \sin kr}{2 \pi^2r}
\frac{K_0\(k \l\)}{k\((1+\beta)K_1\(k\l\)+\beta
k \l
K_0\(k \l\)\)}. \label{VKK1}
\ea
We can either evaluate this integral numerically using some
regularization scheme, or use the following property
\ba
\frac{1}{k}\frac{\(1-\beta\)^2}{
\(1+\beta\)}\frac{K_0(kl)}{(1+\beta)K_1(kl)+kl\beta
K_0(kl)}=  \int_0^{\infty}\d m
\frac{4}{\pi^2 m l\(m^2+k^2\)} \frac{1}{\tilde J_\beta(ml)^2+\tilde Y_\beta(ml)^2},
\ea
where $Y_n$ and $J_\alpha$ are the Bessel functions and we use the
notation:
\ba
\tilde Z_\beta (x)=Z_1(x)+\frac{\beta}{1-\beta} \, x Z_2(x).
\ea
Using this relation, we therefore have:
\ba
V_{KK}&=&\frac{1}{2\pi^2r(1-\beta)}\int \d m\ \frac{4}{\pi^2 m l}
\frac{1}{\tilde J_\beta(ml)^2+\tilde Y_\beta(ml)^2}
\int \d k \frac{k \sin kr}{m^2+k^2} \notag\\
&=& \frac{1}{4\pi r(1-\beta)}\int \d m\ e^{-m r} \frac{4}{\pi^2 m l} \frac{1}{\tilde J_\beta(ml)^2+\tilde
Y_\beta(ml)^2}\notag\\
&\simeq & \frac{1}{4\pi r(1-\beta)}\int \d m\ e^{-m r} m l
\(\frac{1-\beta}{1+\beta}\)^2\left[1-m^2\l^2\(\frac 1 2
-\frac{1-\beta}{1+\beta}\(\gamma-\log\frac{2}{m
\l}\)\)+\cdots\right]\notag\\
&\simeq & \frac{1}{4\pi r}
\frac{1-\beta}{\(1+\beta\)^2}\left[\frac{\l}{r^2}+\frac{2\l^3}{r^4}
\( \frac{4-7\beta}{1+\beta}-3\frac{1-\beta}{1+\beta}\log\frac{2
r}{l} \)
\cdots\right].\label{VKK}
\ea

\end{widetext}

\end{document}